# Big Data Lab
- Databases and Information Systems (DBIS) -

GOETHE UNIVERSITÄT FRANKFURT AM MAIN

# Benchmarking DataStax Enterprise/Cassandra with HiBench

*Technical Report No. 2014-2*
*December 16, 2014*

Todor Ivanov, Raik Niemann, Sead Izberovic, Marten Rosselli,
Karsten Tolle, Roberto V. Zicari

Frankfurt Big Data Laboratory
Chair for Databases and Information Systems
Institute for Informatics and Mathematics
Goethe University Frankfurt
Robert-Mayer-Str. 10,
60325, Bockenheim
Frankfurt am Main, Germany

*www.bigdata.uni-frankfurt.de*

# Table of Contents

# 1. Introduction

This report evaluates the new analytical capabilities of DataStax Enterprise (DSE) [1] through the use of standard Hadoop workloads. In particular, we run experiments with CPU and I/O bound micro-benchmarks as well as OLAP-style analytical query workloads. The performed tests should show that DSE is capable of successfully executing Hadoop applications without the need to adapt them for the underlying Cassandra distributed storage system [2]. Due to the Cassandra File System (CFS) [3], which supports the Hadoop Distributed File System API, Hadoop stack applications should seamlessly run in DSE.

The report is structured as follows: Section 2 provides a brief description of the technologies involved in our study. An overview of our used hardware and software components of the experimental environment is given in Section 3. Our benchmark methodology is defined in Section 4. The performed experiments together with the evaluation of the results are presented in Section 5. Finally, Section 6 concludes with lessons learned.

# 2. Background

**Big Data** has emerged as a new term not only in IT, but also in numerous other industries such as healthcare, manufacturing, transportation, retail and public sector administration [4][5], where it quickly became relevant. There is still no single definition which adequately describes all Big Data aspects [6], but the "*V*" characteristics (*Volume*, *Variety*, *Velocity*, *Veracity* and more) are among the widely used one. Exactly these new Big Data characteristics challenge the capabilities of the traditional data management and analytical systems [6][7]. These challenges also motivate the researchers and industry to develop new types of systems such as Hadoop and NoSQL databases [8].

**Apache Hadoop** [9] is a software framework for distributed storing and processing of large data sets across clusters of computers using the map and reduce programming model. The architecture allows scaling up from a single server to thousands of machines. At the same time Hadoop delivers high-availability by detecting and handling failures at the application layer. The use of data replication guarantees the data reliability and fast access. The core Hadoop components are the Hadoop Distributed File System (HDFS) [10][11] and the MapReduce framework [12].

HDFS has master/slave architecture with a *NameNode* as a master and multiple *DataNodes* as slaves. The *NameNode* is responsible for the storing and managing of all file structures, metadata, transactional operations and logs of the file system. The *DataNodes* store the actual data in the form of files. Each file is split into blocks of a preconfigured size. Every block is copied and stored on multiple *DataNodes*. The number of block copies depends on the *Replication Factor*.

MapReduce is a software framework, that provides general programming interfaces for writing applications that process vast amounts of data in parallel, using a distributed file system, running on the cluster nodes. The MapReduce unit of work is called *job* and consists of input data and a MapReduce program. Each job is divided into *map* and *reduce* tasks. The map task takes a split, which is a part of the input data, and processes it according to the user-defined map function from the MapReduce program. The reduce task gathers the output data of the map tasks and merges them according to the user-defined reduce function. The number of reducers is specified by the user and does not depend on input splits or number of map tasks. The parallel application execution is achieved by running map tasks on each node to process the local data and then send the result to a reduce task which produces the final output.

Hadoop implements the MapReduce model by using two types of processes – *JobTracker* and *TaskTracker*. The *JobTracker* coordinates all jobs in Hadoop and schedules tasks to the *TaskTrackers* on every cluster node. The *TaskTracker* runs tasks assigned by the *JobTracker*. Multiple other applications were developed on top of the Hadoop core components, also known as the Hadoop ecosystem, to make it more ease to use and applicable to variety of industries. Example for such applications are Hive [13], Pig [14], Mahout [15], HBase [16], Sqoop [17] and many more.

**Apache Cassandra** [18][2] is a widely used NoSQL storage system. It has a peer-to-peer distributed ring architecture which can scale to thousands of nodes, communicating with each other via gossip protocol. This makes it capable of storing large data sets, replicated between multiple nodes, with no single point of failure. Cassandra is a key-value store that supports very simple data model with dynamic control over data layout and format [2]. The key is an index in a multi-dimensional map, which represents a Cassandra table, and the value is structured in column family object. Cassandra has a flexible schema and comes with its own query language called Cassandra Query Language (CQL)[19].

**DataStax Enterprise (DSE)** [1] includes production certified version of Apache Cassandra with extended features such as in-memory computing capabilities, advanced security, automatic management services as well as analytical and enterprise search on top of the distributed data. DSE also includes the OpsCenter tool, provided for visual management and monitoring of DSE clusters.

**Cassandra File System (CFS)** [3][20] is an HDFS compatible file system that was built on top of Cassandra to enable running Hadoop applications, without any modification, in DSE. CFS is implemented as a keyspace with two column families. The *inode* column family replaces the HDFS NameNode daemon that tracks each file metadata and block locations. The HDFS DataNode daemon is replaced by the *sblocks* column family that stores the file blocks with actual data. By doing this, the HDFS services are entirely substituted by CFS, removing the single point of failure in the Hadoop NameNode and providing Cassandra with support for large files.

## 3. Setup and Configuration

### 3.1. Hardware

The experiments were performed using Fujitsu BX 620 S3 blade center. With the exception of blade node B1, which is primarily used for administrative tasks of the blade center itself, all the other 8 blade center nodes (nodes B2 to B9) were used. The technical components of blade center B's nodes are listed in Table 1.

| Node | CPU (cores) | Main memory | Hard disk |
|---|---|---|---|
| B1 | 2x AMD Opteron 890 (8) | 32 GByte | 2x 300 GByte as RAID-0 |
| B2 | 2x AMD Opteron 870 (4) | 16 GByte | 2x 143 GByte as RAID-0 |
| B3 | 2x AMD Opteron 870 (4) | 16 GByte | 2x 143 GByte as RAID-0 |
| B4 | 2x AMD Opteron 870 (4) | 16 GByte | 2x 143 GByte as RAID-0 |
| B5 | 2x AMD Opteron 870 (4) | 16 GByte | 2x 143 GByte as RAID-0 |
| B6 | 2x AMD Opteron 870 (4) | 16 GByte | 2x 143 GByte as RAID-0 |
| B7 | 2x AMD Opteron 870 (4) | 16 GByte | 2x 143 GByte as RAID-0 |
| B8 | 2x AMD Opteron 870 (4) | 16 GByte | 2x 143 GByte as RAID-0 |
| B9 | 2x AMD Opteron 870 (4) | 16 GByte | 2x 143 GByte as RAID-0 |

Table 1: Technical Components of Blade Center B

In order to be able to use the maximum of available file storage space, each two hard disks available in the utilized blade nodes were combined into a RAID-0 disk array, enabling nearly 280 GByte of raw file storage space per node.
Each node of blade center B has one 1-GBit internal network adapters that is directly attached to the backplane of the blade center. The backplane acts like a network switch for the blades.

### 3.2. Software

This section briefly describes the software components used in the system under test, as listed in Table 2.

| Software | Version |
|---|---|
| Ubuntu Server 64 Bit | 12.04 LTS |
| Java Runtime Environment | Oracle 1.7.0 60-b19 |
| DataStax Enterprise Sever | 4.0.2 |
| Intel HiBench Benchmark Suite | Adapted Version 2.2 |

Table 2: Software Stack of the System under Test

With the exception of blade node B1, Ubuntu Server 12.04 LTS (64 bit) was installed on all nodes of blade center B. On every node, the two available hard disks were combined into a RAID-0 native disk array and mounted as one logical volume resulting in about 280 GB of storage space. This was done using the logic volume manager tool for Linux kernel, called LVM2 [21]. Three logical volume partitions were created as listed in Table 3. Partition "root" and "data" were formatted using the ext4 file system which resulted in a lower usable partition size compared to the raw size.

| Volume Name | Raw/useable size in GB | Description |
|---|---|---|
| *root* | 20 / 18 | Operating System (Ubuntu Server) |
| *swap* | 8 / 8 | Swap Space |
| *data* | 252 / 246 | Experimental Software (DataStax Enterprise) |

Table 3: Logical Volume Partitions

Version 1.7.0 60-b19 of the Java Runtime Environment (JRE) from Oracle was installed on all nodes as prerequisite for DataStax Enterprise. We installed the OpsCenter [22] tool on one of the nodes and configured it to listen to port 8888 for incoming connections. Using the OpsCenter's web interface we created a Cassandra cluster. The tool takes as parameter the IP addresses of all nodes and automatically performs the cluster setup process by installing the necessary DataStax Enterprise packages and services. It is possible to add new nodes by specifying a pre-generated token value. This value determines the range of the dataset keys that this node will hold. Rebalancing of the key distribution within the cluster is recommended after adding a new node.

In order to enable the Hadoop services (MapReduce, Hive, Pig, Mahout etc.) in DSE, the value of the setting *HADOOP_ENABLED* in the configuration file */etc/default/dse* was changed from 0 to 1. Afterwards the cluster services *dse*, *datastax-agent* and *opscenterd* should be restarted, so that the new settings take effect and the Hadoop components are started.

The default *Replication Factor* of the files in CFS is 1 (keep one copy per file), which is not appropriate for production environments where the data protection and fault tolerance should be guaranteed. Therefore, we changed the replication factor of the *cfs* and *cfs_archive* keyspaces from 1 to 3 (keeping three copies per file) as described in the DSE documentation [23].

All DSE default and adjusted configuration parameters, together with short description, are listed in Table 18 and respectively Table 19 in the Appendix.

### 3.3. HiBench Benchmark Suite

The HiBench [24] benchmark suite was develop by Intel to stress test Hadoop systems. It contains 10 different workloads divided in 4 categories:

1. **Micro Benchmarks** (Sort, WordCount, TeraSort, Enhanced DFSIO)
2. **Web Search** (Nutch Indexing, PageRank)
3. **Machine Learning** (Bayesian Classification, K-means Clustering)
4. **Analytical Queries** (Hive Join, Hive Aggregation)

Each workload has its own specific parameters which will be described later together with a brief description of the workload. Every benchmark reports two metrics: time in seconds and throughput in bytes per second.

In order to run HiBench on the DataStax Enterise server, we had to slightly modify the shell scripts provided in the benchmark suite. This involved prepending the "dse" prefix in front of all Hadoop commands, e.g. `dse hadoop` instead of `hadoop`. Additionally, the HiBench configuration paths were adjusted to run with the Hadoop components in DataStax Enterprise as listed in Table 4. The modified HiBench code is available online [25].

| File | Modification |
|---|---|
| <HiBench-home>/bin/hibench-config.sh | Set variables:<br>*HADOOP_EXECUTABLE="dse hadoop"*<br>*HADOOP_CONF_DIR=/etc/dse/hadoop*<br>*HADOOP_EXAMPLES_JAR=/usr/share/dse/hadoop/lib/hadoop-examples\*.jar*<br>*MAHOUT_LOCAL=/usr/share/dse/mahout* |

Table 4: HiBench Configuration

The <HiBench-home>/bin/run-all.sh script is the main start script, which triggers the execution of all workloads specified in the list contained in the <HiBench-home>/conf/benchmark.lst file. The benchmarks that should not be executed have to be commented out using "#". After each workload execution the results (time and throughtput) are logged in the hibench.report file (in the <HiBench-home> directory).

## 4. Benchmarking Methodology

In this section we describe our benchmarking methodology that we defined and used throughout all experiments. The major motivation behind it was to ensure the comparability between the measured results.

We started by selecting 3 out of the 10 HiBench workloads as listed in Table 5. Our goal was to have representative workloads for CPU bound, I/O bound and mixed CPU and I/O bound workloads.

| Workload | Data structure | CPU usage | IO (read) | IO (write) |
|---|---|---|---|---|
| WordCount | unstructured | high | low | low |
| Enhanced DFSIO | unstructured | low | high | high |
| HiveBench | Structured | high | high | high |

Table 5: Selected HiBench Workload Characteristics

To ensure the accurate performance measurement, each experiment was repeated 3 times and the average value was taken as representative result. Additionally, we report the standard deviation between the measured values in order to prove that the 3 repetitions yield a representative value.

Another rule that we followed during the experiments was to leave approximately 25% of the total storage space that was assigned for the Cassandra cluster to be used for temporary data. In other words, the Cassandra cluster was instructed to use 1476 GByte of the maximum of $8 \times 246$ GByte = 1968 GByte.

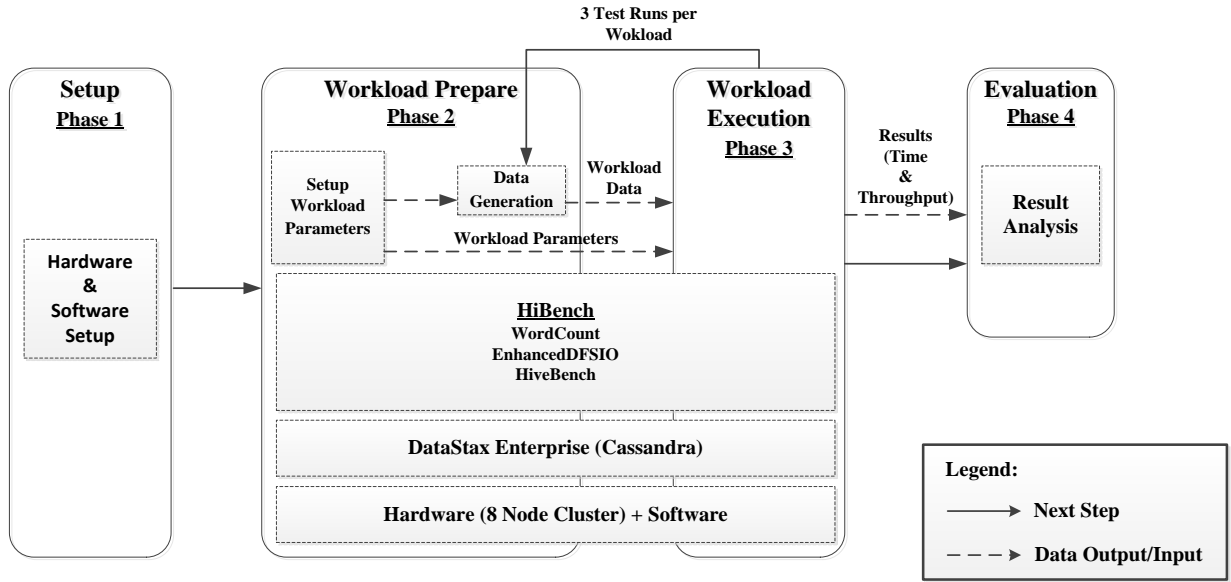

Figure 1: Benchmarking Process Diagram

Figure 1 briefly illustrates the different phases in our experimental methodology. In the initial Phase 1, all software components (OS, Java, DSE and HiBench) are installed and configured. Next in Phase 2, called Workload Prepare, are defined all workload parameters and is generated the test data. The generated data together with the defined parameters are then used as input to execute the workload in Phase 3. As already mentioned each experiment was repeated 3 times to ensure the representativeness of the results, which means that the data generation from Phase 2 and the Workload Execution (Phase 3) were run 3 consecutive times. Before each workload experiment in the Workload Prepare (Phase 2), the existing data is deleted and new one is generated. In Phase 3, HiBench reports two types of results: *Duration* (in seconds) and *Throughput* (in MB per second). The *Throughput* is calculated by dividing the input data size through the *Duration*.

These results are then analyzed in Phase 4, called Evaluation, and presented graphically in the next section of our report.

## 5. Experimental Results

This section describes each of the three HiBench workloads and presents the experimental results together with our evaluation. The results are provided in tables, which consist of multiple columns with the following data:

- **Data Size (GB):** size of the input data in gigabytes
- **Time (Sec):** workload execution duration time in seconds
- **σ (Sec):** standard deviation of **Time (Sec)** in seconds
- **σ (%):** standard deviation of **Time (Sec)** in percent
- **Data Δ (%):** difference of **Data Size (GB)** to a given data baseline in percent
- **Time Δ (%):** difference of **Time (Sec)** to a given time baseline in percent

### 5.1. WordCount

**WordCount** is a MapReduce job which calculates the number of occurrences of each word in a text file. The input text data is generated by the RandomTextWriter program which is also part of the standard Hadoop distributions.

#### 5.1.1. Preparation

The WordCount workload takes 3 parameters listed in Table 6. The DATASIZE parameter is relevant only for the data generation.

| Parameter | Description |
|---|---|
| NUM_MAPS | Number of mappers |
| NUM_REDS | Number of reducers |
| Relevant for the data generator | |
| DATASIZE | Size of the text file to generate |

Table 6: WordCount Parameters

#### 5.1.2. Results and Evaluation

In order to find the optimal numbers of mappers and reducers for the WordCount workload on our hardware configuration, we performed some tests as described in the following subsection 5.1.2.1. In these tests the DATASIZE parameter was fixed to 240 GB. The result of this test was used to perform experiments with different DATASIZE parameters as described in subsection 5.1.2.2.

##### 5.1.2.1. Optimal Number of Mappers and Reducers

The goal of this experiment was to find optimal values for the mappers and reducers for our hardware configuration. The best practice [26] is to configure 1.5 mapper and reducer tasks for each physical CPU core. In our case, we have 4 physical CPU cores, which results in 6 tasks in total. On the other hand, the rule of thumb states [26] that roughly two thirds of the slots should be allocated to map tasks and the remaining one third as reduce tasks. In our case, we should have 4 map and 2 reduce tasks.

Therefore, the input data size was fixed to 30GB per node (parameter DATASIZE) and the number of mappers and reducers (parameter NUM_MAPS and NUM_REDS per node) was chosen following the best practices. In order to find the optimal configuration of map and reduce slots for our setup, all experiments listed in Table 7 were performed.

| Total Data Size (GB) | DATASIZE per Node (Bytes) | NUM_MAPS | NUM_REDS |
|---|---|---|---|
| 240 | 32212254720 | 3 | 1 |
| 240 | 32212254720 | 4 | 2 |
| 240 | 32212254720 | 6 | 2 |
| 240 | 32212254720 | 12 | 4 |

Table 7: WordCount Map/Reduce Experiments

Figure 2 shows the job processing times (in seconds) and the throughput (MBs per second) for each of the four configurations. *The lower values in the Time graph represent faster completion times, respectively the higher values in the Throughput graph account for better performance.* Clearly the configuration with 4 map and 2 reduce tasks achieves the best time and confirms the best practices. The throughput for all of the experiments is almost identical which can be explained with the fact that the WordCount workload is very CPU intensive, with light disk and network usage. Therefore, we configure our further experiments with 4 map and 2 reduce tasks.

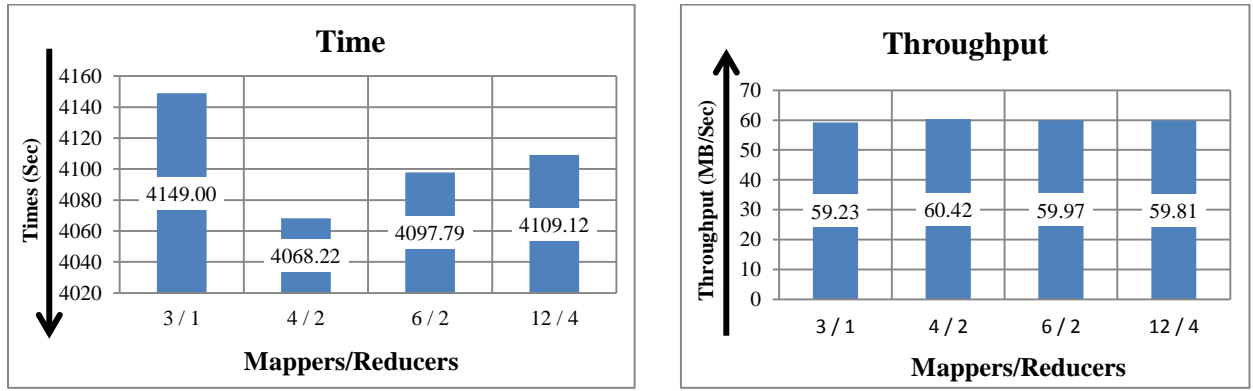

Figure 2: WordCount - Different Mappers/Reducers Experiments

### 5.1.2.2. Processing Different Data Sizes

The motivation behind this experiment is to observe how the performance of CPU bound applications changes with the increase of the input data size. For all the experiments the number of mappers and reducers (parameter NUM_MAPS and NUM_REDS) were fixed to 4 and 2. Experiments were performed for three different input data sizes (parameter DATASIZE) 240, 340 and 440 GBs, also shown in Table 8.

| Total Data Size (GB) | DATASIZE per Node (Bytes) | NUM_MAPS | NUM_REDS |
|---|---|---|---|
| 240 | 32212254720 | 4 | 2 |
| 340 | 45634027520 | 4 | 2 |
| 440 | 59055800320 | 4 | 2 |

Table 8: Variable Data Size Parameters

Figure 3 depicts the time (in seconds) and throughput (in MBs per second) for the three different test scenarios. As expected the increase in data size leads to increased processing time, but the throughput stays constant as the workload is heavily CPU-bound.

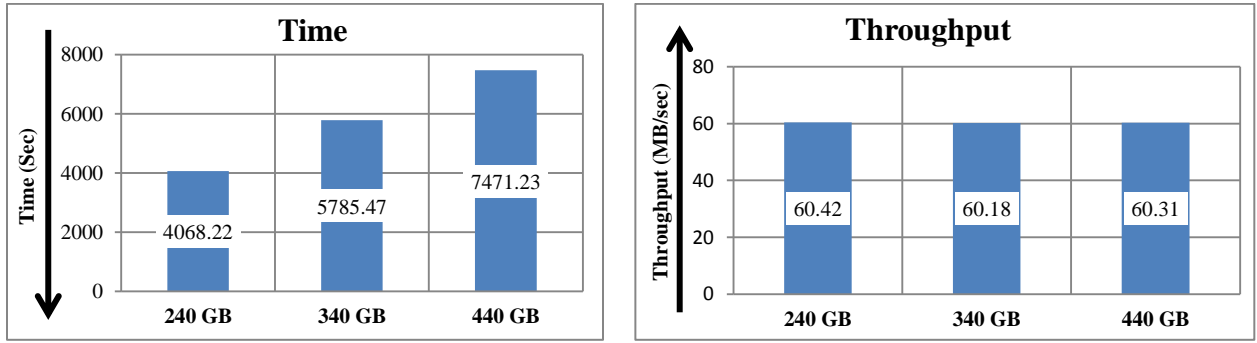

Figure 3: WordCount – Processing Different Data Sizes

Table 9 summarizes the times together with the standard deviation (**σ (Sec)** and **σ (%)**) and the **Time Δ** (in %) differences between the baseline experiment (240GB) and the other two (340GB and 440GB). It is interesting to observe that increasing the data size with 100GB (+42% more) increases the total processing time with around 42%. Respectively, doubling the data increase to 83% from 240GB to 440GB increases the processing time with almost 84%.

| Data Size (GB) | Time (Sec) | σ (Sec) | σ (%) | Data Δ (%) | Time Δ (%) |
|---|---|---|---|---|---|
| 240 | 4068.22 | 57.18 | 1.41 | baseline | baseline |
| 340 | 5785.47 | 17.36 | 0.30 | +41.67 | +42.21 |
| 440 | 7471.23 | 34.55 | 0.46 | +83.33 | +83.65 |

Table 9: WordCount Results

Figure 4 illustrates that the system scales nearly linearly with the increase of the data size. We observe that the data points nearly match the stepwise linear scaling, which is represented by the green line.

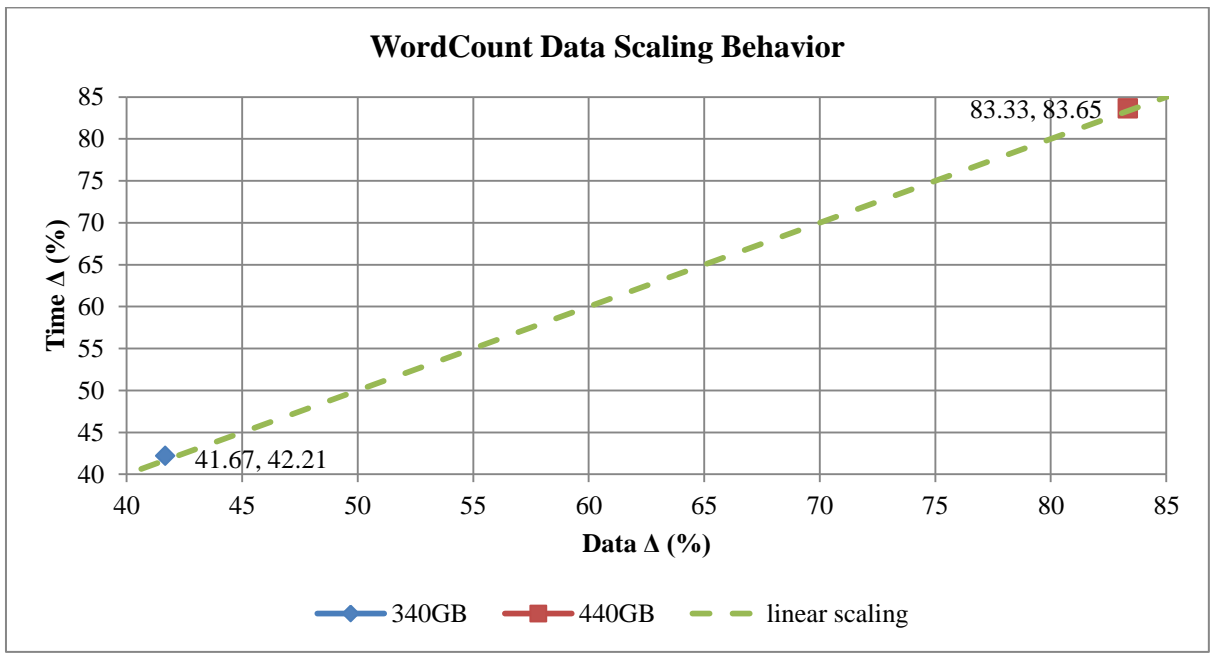

Figure 4: WordCount Data Scaling Behavior

In summary, our experiments showed that DSE is capable of running compute-intensive MapReduce applications, achieving stepwise linear performance with the growing data sizes.

## 5.2. Enhanced DFSIO

**TestDFSIO** [27] is a HDFS benchmark included in Hadoop distributions. It is designed to stress test the storage I/O (read and write) capabilities of a cluster. In this way performance bottlenecks in the network, hardware, OS or Hadoop setup can be found and fixed. The benchmark consists of two parts: TestDFSIO-write and TestDFSIO-read. The write program starts multiple map tasks with each task writing a separate file to HDFS. The read program starts multiple map tasks with each task sequentially reading the previously written files and measuring the *file size* and the task *execution time*. The benchmark uses a single reduce task to measure and compute two performance metrics for each map task: *Average I/O Rate* and *Throughput*. Respectively, Equation 1 and Equation 2 illustrate how the two metrics are calculated with **N** as the total number of map tasks and the index **i** (0< i < N), identifying the individual tasks.

Equation 1: *Average I/O Rate*

$$\textit{Average I/O rate (N)} = \frac{\sum_{i=1}^{N} rate(i)}{N} = \frac{\sum_{i=1}^{N} \frac{file\ size(i)}{time(i)}}{N}$$

Equation 2: *Throughput*

$$\textit{Throughput (N)} = \frac{\sum_{i=1}^{N} file\ size(i)}{\sum_{i=1}^{N} time(i)}$$

**EnhancedDFSIO** is an extension of the DFSIO benchmark developed specifically for HiBench [24]. The original TestDFSIO benchmark reports the average I/O rate and throughput for a single map task, which is not representative in cases when there are delayed or re-tried map tasks. EnhancedDFSIO addresses the problem by computing the aggregated I/O bandwidth. This is done by sampling the number of bytes read/written at fixed time intervals in the format of (map id, timestamp, total bytes read/written). Aggregating all sample points for each map tasks allows plotting the exact map task throughput as linearly interpolated curve. The curve consists of a warm-up phase and a cool-down phase, where the map tasks are started and shut down, respectively. In between is the steady phase, which is defined by a specified percentage (default is 50%, but can be configured) of map tasks. When the number of concurrent map tasks at a time slot is above the specified percentage, the slot is considered to be in the steady phase. The EnhancedDFSIO aggregated throughput metric is calculated by averaging the value of each time slot in the steady phase.

### 5.2.1. Preparation

The EnhancedDFSIO takes four input configuration parameters as described in Table 10.

| Parameter | Description |
|---|---|
| RD_FILE_SIZE | Size of a file to read in MB |
| RD_NUM_OF_FILES | Number of files to read |
| WT_FILE_SIZE | Size of a file to write in MB |
| WT_NUM_OF_FILES | Number of files to write |

Table 10: EnhancedDFSIO Parameters

For the EnhancedDFSIO benchmark, the file sizes (parameters RD_FILE_SIZE and WT_FILE_SIZE), which the workload should read and write, were fixed to 400MB. In the same time, the number of files (parameters RD_NUM_OF_FILES and RD_NUM_OF_FILES) were fixed to be 615, 871 and 1127 to operate on a data set with data sizes of 240, 340 and 440 GB. The total data size is the product of multiplying the specific file size with the number of files to be read/written. Three experiments were executed as listed in Table 11.

| Data Size (GB) | RD_FILE_SIZE | RD_NUM_OF_FILES | WT_FILE_SIZE | WT_NUM_OF_FILES |
|---|---|---|---|---|
| 240 | 400 | 615 | 400 | 615 |
| 340 | 400 | 871 | 400 | 871 |
| 440 | 400 | 1127 | 400 | 1127 |

Table 11: EnhancedDFSIO Experiments

### 5.2.2. Results and Evaluation

Figure 5 depicts read and write times for the different input data sizes. It is interesting to observe that the difference between reading and writing times is very small, as reported in Table 12. Comparing the times for 240GB, we observe that the **Read/Write Δ** is around 6% and gradually decreases to around 3% for 440GB data size. The reason for this writing times is the architecture of Cassandra, which morphs all writes to disk into sequential writes [2] resulting in maximum write throughput as we can see on Figure 6.

| Data Size (GB) | Read Times (Sec) | Write Times (Sec) | Read/Write Δ (%) |
|---|---|---|---|
| 240 | 915.61 | 973.90 | 6.37 |
| 340 | 1405.67 | 1477.35 | 5.10 |
| 440 | 2050.84 | 2110.06 | 2.89 |

Table 12: EnhancedDFSIO Read/Write Δ

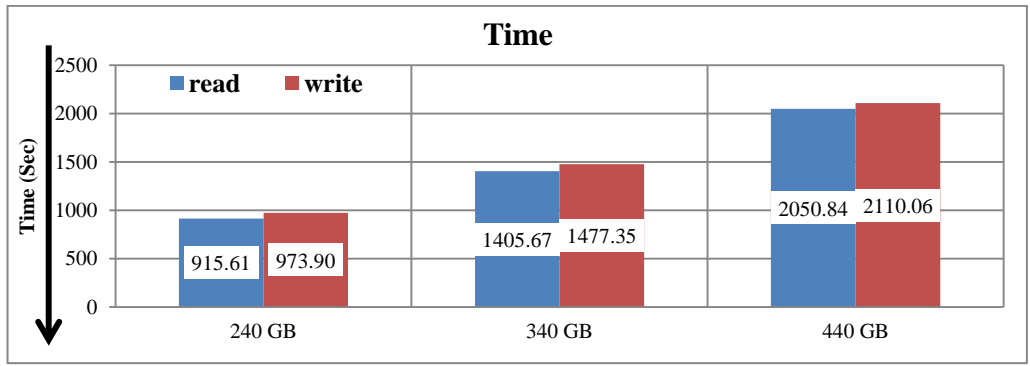

Figure 5: EnhancedDFSIO – Time

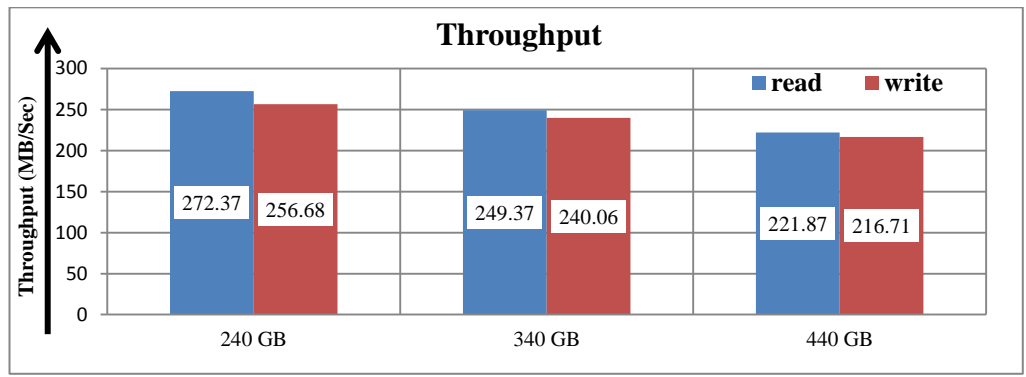

Figure 6: EnhancedDFSIO – Throughput

In a related work, Dede *et al.* [28] report similar results in their experiments with 8 node Cassandra cluster, where reading has drastically improved when increasing the number of nodes. In the same time, increasing the data size from 16 to 32 million records has also improved the reading performance, thus decreasing the gap between reading and writing.

Table 13 summarizes the processing times together with the standard deviations (**σ (Sec)** and **σ (%)**) obtained from the 3 consequent test runs. There is also a column representing the **Time Δ** (in %) between the baseline data size (240GB) and the other two input data sizes. It is interesting that the deltas for both reading and writing operations are very similar. Increasing the data size with 100GB takes around 52-54% more time for both operations. Similarly, doubling the data size yields around 117-124% more processing time for both operations.

| Data Size (GB) | Test | Time (Sec) | σ (Sec) | σ (%) | Data Δ (%) | Time Δ (%) |
|---|---|---|---|---|---|---|
| 240 | Read | 915.61 | 95.18 | 10.39 | baseline | baseline |
| 340 | Read | 1405.67 | 20.16 | 1.43 | +41.67 | +53.52 |
| 440 | Read | 2050.84 | 147.22 | 7.18 | +83.33 | +123.99 |
|  |  |  |  |  |  |  |
| 240 | Write | 973.90 | 75.00 | 7.70 | baseline | baseline |
| 340 | Write | 1477.35 | 135.74 | 9.19 | +41.67 | +51.69 |
| 440 | Write | 2110.06 | 121.88 | 5.78 | +83.33 | +116.66 |

Table 13: DFSIOEnh Results

Figure 7 depicts the *DFSIO-read* and *DFSIO-write* results for our experimental data points with respect to the linearly projected scaling line in green. We observe that in both cases the data points lie higher than the projected line, which means that the system does not scale linearly with the increase of the data size.

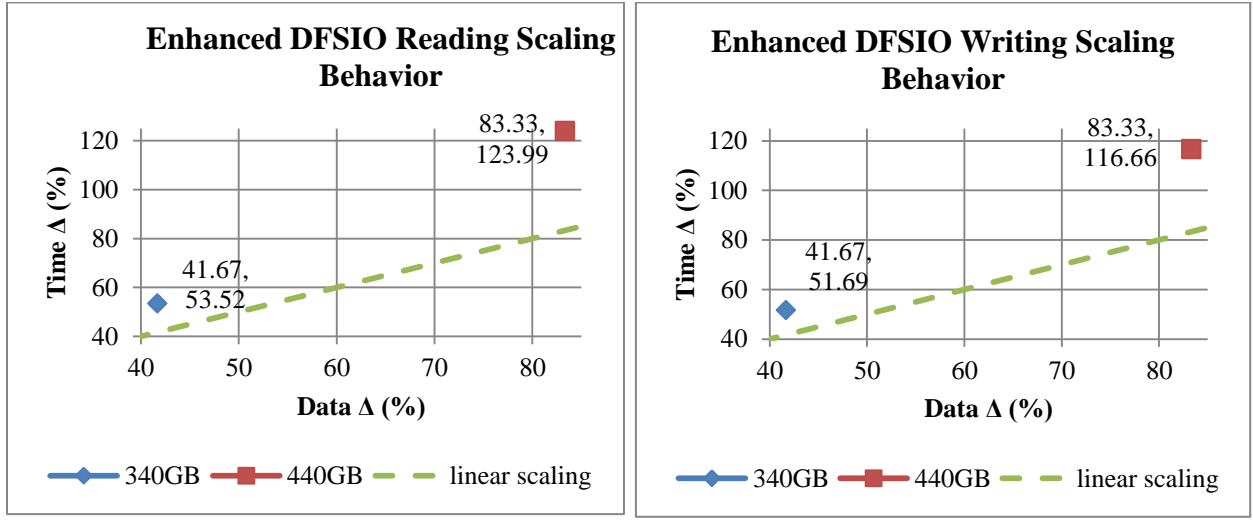

Figure 7: EnhancedDFSIO - Scaling Behavior

In summary, our experiments show that DSE provides good write capabilities, which are slightly slower than the read operations. Interestingly, we observed that increasing the input data size decreases the difference between read and write operations.

### 5.3. HiveBench

The OLAP-style analytical queries, called **HiveBench,** are adapted from the Pavlo's Benchmark [29] and have the goal to test the performance of Hive [13], running on top of MapReduce. Table 14 shows the exact SQL code for the Hive Join and Aggregation queries included in the benchmark. Additionally, there are two tables *Rankings* (default size of 1GB) and *UserVisits* (default size of 20GB), also listed in Table 14. Both tables are used in the join query, whereas the aggregation uses only the *UserVisits* table.

| Application | Query |
|---|---|
| Hive Join | `SELECT sourceIP, sum(adRevenue) as totalRevenue, avg(pageRank) FROM rankings R JOIN (SELECT sourceIP, destURL, adRevenue FROM uservisits UV WHERE (datediff(UV.visitDate, '1999-01-01')>=0 AND datediff(UV.visitDate, '2000-01-01')<=0)) NUV ON (R.pageURL = NUV.destURL) GROUP BY sourceIP ORDER BY totalRevenue DESC limit 1;` |
| Hive Aggregation | `SELECT sourceIP, SUM(adRevenue) FROM uservisits GROUP BY sourceIP;` |
| Table Rankings | `CREATE TABLE rankings (pageURL STRING, pageRank INT, avgDuration INT);` |
| Table UserVisits | `CREATE EXTERNAL TABLE uservisits (sourceIP STRING,destURL STRING,visitDate STRING,adRevenue DOUBLE,userAgent STRING,countryCode STRING,languageCode STRING,searchWord STRING,duration INT );` |

Table 14: HiBench Analytical Queries; Adopted from [30]

The implementation of the join query in Hive is not as trivial as the SQL code. It consists of three separate phases, which are actually three MapReduce programs executed sequentially, and a temporary table. The aggregation query performs parallel analytics on the *UserVisits* table, which in Hadoop are also implemented as multiple map, reduce and combine tasks. More detailed description of the queries is provided in [29].

### 5.3.1. Preparation

HiveBench benchmark has four parameters listed in Table 15 which are also relevant for the data generation.

| Parameter | Description |
|---|---|
| NUM_MAPS | Number of mappers |
| NUM_REDS | Number of reducers |
| PAGES | Number of pages |
| USERVISITS | Number of user visits |

Table 15: HiveBench Parameters

For the HiveBench workloads the data sizes are defined in millions of pages and approximately estimated in GBs for the two parameters PAGES and USERVISITS as follows:

$$PAGES = 12.000.000 \approx 1 \text{ GB}$$
$$USERVISITS = 100.000.000 \approx 20 \text{ GBs}$$

The other two parameters NUM_MAPS and NUM_REDS are relevant only for the data generation stage and does not influence the workload performance. The exact test values for all test parameters are listed in Table 16.

| Total Data Size (GB) | PAGES | USERVISITS | NUM_MAPS | NUM_REDS |
|---|---|---|---|---|
| 240 | 132 000 000 | 1 100 000 000 | 220 | 110 |
| 340 | 204 000 000 | 1 700 000 000 | 340 | 170 |
| 440 | 264 000 000 | 2 200 000 000 | 440 | 220 |

Table 16: HiveBench Experiments

### 5.3.2. Results and Evaluation

Figure 8 shows the times (in seconds) for both hive-aggregation and hive-join workloads. *The lower values in the Time graphic represent faster completion times, respectively the higher values in the Throughput graphic account for better performance.* It is interesting to observe that for all the three experiments hive-aggregation performs slower than hive-join.

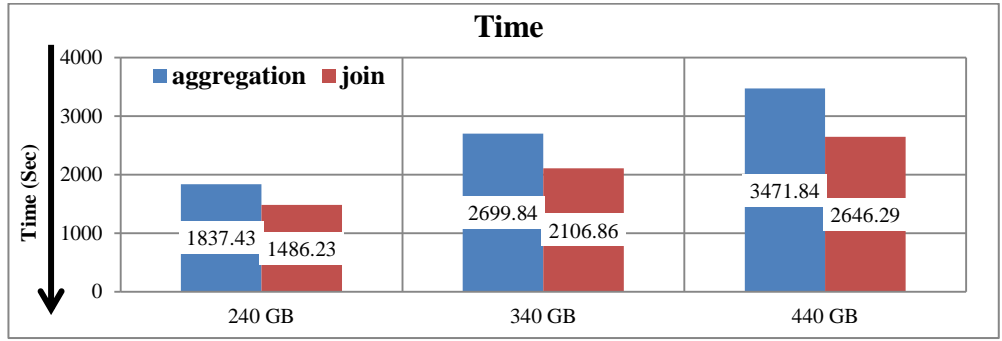

Figure 8: HiveBench - Time

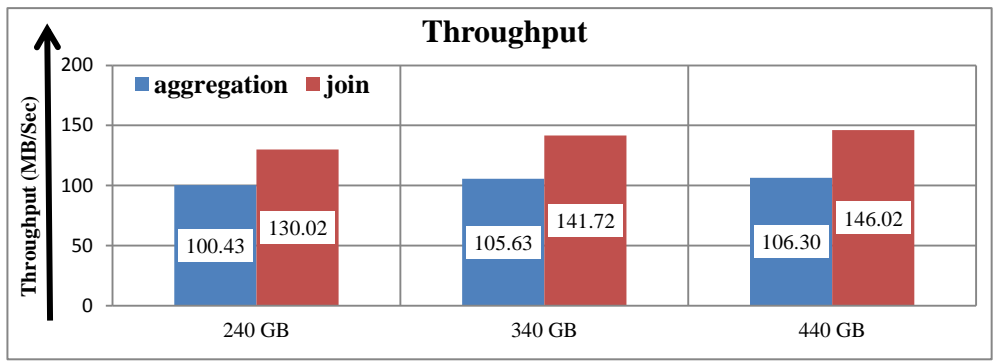

Figure 9: HiveBench - Throughput

Figure 9 depicts the throughput (in MBs per second) for both workloads and clearly shows that the hive-aggregation achieves a smaller throughput than hive-join for all the three experiments. Additionally we observe that with the increase in data size the throughput of hive-aggregation slightly increases and remains constant for the last two experiments. This is not the case with hive-join where we observe greater increase in throughput.

Table 17 summarizes the times (in seconds) for both hive-aggregation and hive-join together with the standard deviations (**σ (Sec)** and **σ (%)**) for the three experiments. Increasing the data size with 100GB (+42%) increases the processing time for hive-aggregation around 47% and for hive-join around 42%. By increasing the baseline with 200GB (around 83% more) to 440GB data size, the processing time increased around 89% for hive-aggregate and respectively around 78% for hive-join. These numbers clearly show that both hive-aggregate and hive-join queries scale almost linearly within 5% range with the increase of the data sets.

| Total Data Size (GB) | Test | Processed Data (GB) | Time (Sec) | σ (Sec) | σ (%) | Data Δ (%) | Time Δ (%) |
|---|---|---|---|---|---|---|---|
| 240 | Aggregation | 180.20 | 1837.43 | 5.52 | 0.30 | baseline | baseline |
| 340 | Aggregation | 278.50 | 2699.84 | 10.81 | 0.40 | +41.67 | +46.94 |
| 440 | Aggregation | 360.40 | 3471.84 | 20.24 | 0.58 | +83.33 | +88.95 |
|  |  |  |  |  |  |  |  |
| 240 | Join | 188.68 | 1486.23 | 25.56 | 1.72 | baseline | baseline |
| 340 | Join | 291.60 | 2106.86 | 10.75 | 0.51 | +41.67 | +41.76 |
| 440 | Join | 377.35 | 2646.29 | 3.01 | 0.11 | +83.33 | +78.05 |

Table 17: HiveBench Results

Figure 4 illustrates how the system scales with the increase of the data size. We observe that for hive-aggregation the systems perform nearly linear with the data points lying slightly above the projected green line, whereas for hive-join the data point for the 340GB experiment lies nearly on the linearly projected green line. Interestingly, by increasing the data size to 440 GB the data point for hive-join lies below the green line, which means it scales better than linear.

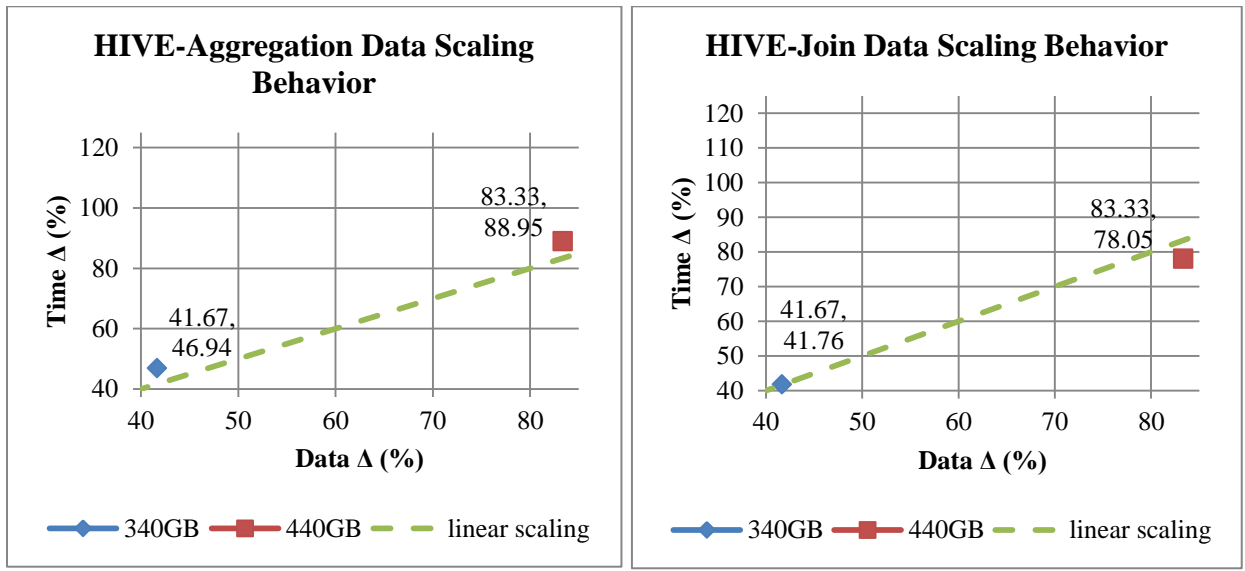

Figure 10: HIVE – Data Scaling Behavior

Overall, our experiments show that OLAP-style analytical queries can be successfully run on top of DSE.

## 6. Lessons Learned

In this report we present results showing that the HiBench Benchmark Suite, developed specifically for Hadoop, can be successfully executed on top of the DataStax Enterprise (DSE). Our experiments stress tested the DSE platform by performing multiple runs with CPU bound, I/O bound and analytic MapReduce workloads. Our results showed:
- CFS can be used instead of or complementary to HDFS as distributed storage layer for the tested data sets (240-440 GB) of both structured and unstructured data.
- For WordCount (see page 7) DSE scales nearly linear with the increase of the data size.
- The Enhanced DFSIO experiments (see page 10) demonstrated that DSE provides good write operations, which are slightly slower than the read operations.
- The HiveBench tests (see page 13) showed that DSE can successfully run OLAP-style analytical queries on the tested data sets (240-440 GB).

# Appendix

| Parameter | Default | Description |
|---|---|---|
| io.seqfile.compress.blocksize | 1048576 | The minimum block size for compression in block compressed SequenceFiles. |
| fs.local.block.size | 67108864 | Default: 64 MB This is the default block size, in bytes, for new files created in the distributed file system. |
| fs.local.subblock.size | 2097152 | Default: 2 MB SubBlock Size |
| io.sort.factor | 12 | This value sets the number of input files that are merged at once by map/reduce tasks. |
| io.sort.mb | 128 | This sets the size of memory buffer used during sort operations. |
| mapred.child.java.opts | -Xmx256M | This parameter is used to pass any Java options to the map/reduce child processes, and we will use this to set the maximum Java heap size for each map/reduce task. |
| mapred.tasktracker.map.tasks.maximum | 3 | These are the maximum number of map/reduce tasks permitted to execute simultaneously per node. |
| mapred.tasktracker.reduce.tasks.maximum | 3 | These are the minimum number of map/reduce tasks permitted to execute simultaneously per node. |
| mapred.job.inode.mode | SYNC | Performance improvement<br>The inode can be saved once per block or once per job and whe the file is closed.<br>The latter will offer fast performance at expense or more risk.<br>Options:<br>SYNC (one save per block) (Default)<br>ONCE (inode is saved at the end) |
| mapred.job.reuse.jvm.num.tasks | 24 | |
| mapred.compress.map.output | true | Compress intermediate files for better performance |
| mapred.map.output.compression.codec | com.datastax.bdp.hadoop.compression.SnappyCodec | |

Table 18: DataStax Enterprise Configuration Parameters

| Parameter | Value | Description |
|---|---|---|
| io.sort.mb | 256 | This sets the size of memory buffer used during sort operations. |
| mapred.map.child.java.opts | -Xmx1G | Larger heap-size for child jvms of maps. |
| mapred.reduce.child.java.opts | -Xmx1G | Larger heap-size for child jvms of reduces. |

Table 19: DataStax Enterprise Adjusted Parameters

# Acknowledgements

This research is supported by the Big Data Lab at the Chair for Databases and Information Systems (DBIS) at the Goethe University Frankfurt and the Institute for Information Systems (IISYS) at Hof University of Applied Sciences.